# SIMULATION OF REACTION-DIFFUSION PROCESSES IN THREE DIMENSIONS USING CUDA

Ferenc Molnár Jr.[1], Ferenc Izsák[2], Róbert Mészáros[3], István Lagzi[3,4*]

[1] Department of Theoretical Physics, Eötvös Loránd University, Budapest, Hungary

[2] Department of Applied Analysis and Computational Mathematics, Eötvös Loránd University, Budapest, Hungary

[3] Department of Meteorology, Eötvös Loránd University, P.O. Box 32, H-1518 Budapest, Hungary

[4] Department of Chemical and Biological Engineering, Northwestern University, Evanston, Illinois, USA

**Abstract**

Numerical solution of reaction-diffusion equations in three dimensions is one of the most challenging applied mathematical problems. Since these simulations are very time consuming, any ideas and strategies aiming at the reduction of CPU time are important topics of research. A general and robust idea is the parallelization of source codes/programs. Recently, the technological development of graphics hardware created a possibility to use desktop video cards to solve numerically intensive problems. We present a powerful parallel computing framework to solve reaction-diffusion equations numerically using the Graphics Processing Units (GPUs) with CUDA. Four different reaction-diffusion problems, (i) diffusion of chemically inert compound, (ii) Turing pattern formation, (iii) phase separation in the wake of a moving diffusion front and (iv) air pollution dispersion were solved, and additionally both the Shared method and the Moving Tiles method were tested. Our results show that parallel implementation achieves typical acceleration values in the order of 5–40 times compared to CPU using a single-threaded implementation on a 2.8 GHz desktop computer.

[*] Corresponding author. Tel.: +36-2090555; fax: +36-1372-2904.

1. **Introduction**

There are many spectacular and fascinating phenomena (Figure 1) in the nature and laboratories [1], which can be described and understood by reaction-diffusion systems (e.g. autocatalytic front propagation [2], chemical waves [3], Turing patterns [4,5], seashell pattern formation [6], Liesegang phenomenon [7], etc.). Generally, reaction-diffusion systems are mathematical models that describe the spatial and temporal variations of concentrations of chemical substances involved in a given process. From the mathematical point of view, the reaction-diffusion system is a set of parabolic partial differential equations (PDEs), and it has a general form:

$$\frac{\partial \vec{c}}{\partial t} = -\nabla \cdot \left(-\mathbf{D}\nabla \vec{c}\right) + \mathbf{R}(\vec{c}), \qquad (1)$$

where $\vec{c} = \left(c_1(t,\mathbf{x}), c_2(t,\mathbf{x}), ..., c_k(t,\mathbf{x})\right)$ denotes the concentration set of the chemical species, **D** is a diagonal matrix consisting of the diffusion coefficients $D_1, D_2,..., D_k$, $\nabla$ denotes the del operator, and **R**, which is usually nonlinear term, represents the chemical reactions. Equation (1) can be rewritten into a more specialized form if the diffusion coefficients do not depend on location (i.e. diffusion processes are isotropic):

$$\frac{\partial \vec{c}}{\partial t} = \mathrm{D}\nabla^2 \vec{c} + \mathbf{R}(\vec{c}), \qquad (2)$$

*E-mail address*: lagzi@vuk.chem.elte.hu (I. Lagzi).

where $\nabla^2$ is the Laplace operator.

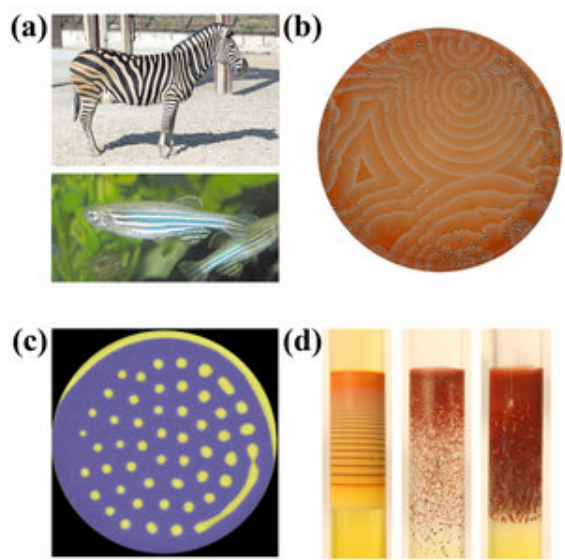

**Figure 1** Reaction-diffusion patterns in animate and inanimate systems. (a) Striped patterns on skin of animals (zebra and zebra fish); (b) chemical waves in a Belousov-Zhabotinsky reaction; (c) Turing pattern in a 2D gel sheet (image courtesy of Dr. István Szalai); (d) precipitation (Liesegang) pattern formation.

In recent years, the technological development of consumer graphics hardware has created a possibility to use desktop video cards to solve numerically intensive problems in various fields of science (chemistry and physics [8–15], astronomy [16–18], medical sciences [19–21], geosciences [22,23], environmental sciences [24–26] and mathematics [27,28]), since their computational capacity far exceeds that of the desktop CPUs [29–31]. Using GPUs (processors of video cards) for general purpose calculations is called GPGPU. Its main advantage is the high cost-effectiveness compared to supercomputers, clusters or GRID systems. Programming GPUs for general computation was a great challenge in the past, but NVIDIA has created a parallel computing architecture called Compute Unified Device Architecture, or CUDA [29], which significantly simplifies the programming. Programs can be written in the well-known C language with some CUDA-specific extensions. The NVIDIA nvcc compiler, a software development kit with utilities, libraries and numerous examples, and also a complete documentation are freely available [32].

There have only been a few trials in the literature to solve various types of PDEs using CUDA environment [27, 33–35]. In this paper we present efficient techniques to utilize GPU computing power using CUDA to solve several reaction-diffusion problems in three spatial dimensions. This new method provides a much more efficient way to perform these simulations than using CPUs of desktop computers.

2. **Numerical implementation of reaction-diffusion systems**

The most convenient and common technique for solving time dependent PDEs is called the method of lines, where "line" refers to the time levels. This approach reduces the set of PDEs in three independent variables to a system of ordinary differential equations (ODEs) in one independent variable, time. The system of ODEs can then be solved as an initial value problem. Usually the grid can be fixed over the computational domain, where the unknown function of physical quantities (here a vector function) is estimated in each time step. In a usual approach, which we follow, the spatial derivatives in the equation are approximated with finite differences at a fixed time. In the reaction-diffusion equations only the Laplacian differential operator is present. Approximation of Laplacian for some function $c$ at any grid point can be performed by calculating a linear combination of the neighbouring grid points using a specific set of coefficients applying Taylor expansion. The following nineteen-point approximation for Laplacian was used for 3D simulations (supposing equidistant gridding in all dimensions):

$$Lap_{i,j,k}^l = \nabla^2 c_{i,j,k}^l = \frac{1}{6h^2} \sum_{p=-1,q=-1,r=-1}^{p=1,q=1,r=1} T_{pqr} c_{i+p,j+q,k+r}^l, \qquad (3)$$

$$T_{p,q,-1} = T_{p,q,1} = \begin{pmatrix} 0 & 1 & 0 \\ 1 & 2 & 1 \\ 0 & 1 & 0 \end{pmatrix}, \quad T_{p,q,0} = \begin{pmatrix} 1 & 2 & 1 \\ 2 & -24 & 2 \\ 1 & 2 & 1 \end{pmatrix} \tag{4}$$

where $c_{i,j,k}$ is the value of $c$ on the grid point $(i, j, k)$, $h$ is the spatial resolution of the grid (grid spacing) in all three dimensions, and $l$ corresponds to the given chemical species involved in reaction-diffusion process. This approximation takes more computational effort than the more commonly used seven-point stencil. While its precision is the same, it provides better isotropy on rectangular grids.

The values at the next time level can be obtained with an explicit or implicit time stepping. We start solving the initial value problem with initial condition $\vec{c}(t=0) = \vec{c}_0$. The simplest numerical integration scheme – forward Euler method – was used, which gives that for any time $t$

$$c_{i,j,k}^l (t + \delta t) = c_{i,j,k}^l (t) + \left[ \mathbf{D} \, Lap_{i,j,k}^l + R\left(c_{i,j,k}^l (t)\right) \right] \delta t, \tag{5}$$

here $\delta t$ is the time step. Although a scale of more powerful time stepping methods are available, our aim is rather to illustrate the power of the computational force provided by CUDA.

## 3. Basics of CUDA

The main concept of CUDA parallel computing model is to operate with tens of thousands of lightweight *threads*, grouped into *thread blocks*. These threads must execute the same function with different parameters. This function, which contains all the computations and runs in parallel in many instances is called the *kernel*. Instances of the kernel are identified by thread and block indices. Threads in the same thread block can synchronize execution with each other, by inserting synchronization points in the kernel, which must be reached by all threads in the block before continuing execution. These threads can also share data during execution. This way several hundred threads in the same block can work cooperatively. Threads of different thread blocks cannot be synchronized and should be considered to run independently.

It is possible to use a small number of threads and/or small number of blocks to execute a kernel, however, it would be very inefficient. This would utilize only a fraction of the computing power of the GPU. Therefore, CUDA is the best suited to those problems that can be divided into many parts, which can be computed independently (in different blocks), and these should be further divided into smaller cooperating pieces (into threads).

There are several types of memory available in CUDA designed for different uses in kernels. If used properly, they can increase the computation performance significantly. The *global memory* is essentially the random access video memory available on the video card. It may be read or written any time at any location by any of the threads, but to achieve high performance access to global memory should be *coalesced*, meaning the threads must follow a specific memory access pattern. More complete (and hardware-revision dependent) description can be found in the Programming Guide [36]. A kernel has access to two cached, read-only memories: the *constant memory* and the *texture memory*. Constant memory may be used to store constants that do not change during kernel execution, and all instances of a kernel use them regardless of thread and block indices. Texture memory may be used efficiently when threads access data with spatial locality in one, two, or three dimensions. It also provides built-in linear interpolation of the data. There is also a parallel data cache available for reading and writing for all the threads of the same thread block called the *shared memory*. It makes the cooperative work of threads in a block possible. It is divided into 16 banks. Kernels should be written in a way to avoid bank conflicts, meaning the threads which are executed physically at the same time should access different banks [36].

Memory management and kernel execution are controlled by CUDA library functions in the *host* code (the one which runs on the CPU). While the kernels are executing on the *device*, the CPU continues to execute host code, so CPU and GPU can work in parallel.

Up till now, many CUDA-capable video cards and other computing devices were produced with different capabilities. All devices have a special version number which indicates the GPU's computational skillset, called *compute capability*. It is important to distinguish devices with compute capability 1.0 and 1.1 from those with compute capability 1.2 and 1.3. The former ones represent the first generation of CUDA devices, based on the G80 GPU, while the latter are based on the more advanced GT200 GPU. Although the basic concepts apply to all CUDA devices, the two major versions have different rules for achieving maximum performance.

## 4. Application

Solving reaction-diffusion equations fits well to the architecture of CUDA. The basic ideas are the following. Concentrations of the species in a simulation can be stored in global memory on the device. We can assign computation of the next time level to a kernel function, assigning threads to compute individual grid points. The rectangular space represented by the grid points can easily be splitted to smaller parts, which can be assigned to blocks. Shared memory within a block utilised in the approximation of the spatial derivatives, because this computation requires data which belongs to neighbouring grid points (and threads). Physical constants and other parameters of a simulation can be stored in the constant memory.

Two very important performance guidelines must be followed to reach maximum performance: accessing (reading or writing) global memory should follow a *coalesced* access pattern, and accesses to the shared memory should be without bank conflicts. CUDA devices from the first generation have very strict rules for achieving coalesced memory access, which result in a computational solution for them and another one for the second generation. However, a solution which is optimized well for the first generation of devices should also run very efficiently on the devices of second generation with minor adjustments to some parameters.

Four different reaction-diffusion problems were chosen and solved using CUDA to illustrate the capability and efficiency of GPU computing. The first one is the pure diffusion, which presents the 'core' mechanism in all reaction-diffusion related problems. The second example is the most famous and well-known Turing pattern formation. Here the diffusion is coupled to nonlinear chemical reactions. This framework can be applied to many reaction-diffusion problems. The third problem describes a phase separation in chemical systems using the Cahn–Hilliard equation. The curiosity of this equation originates from the fact that it contains fourth order spatial derivatives. Our last example is an extension of the pure diffusion transport problem with advection. This arises in many areas, especially in air pollution, where the transport of air pollutants consist of two main transport phenomena (advection – transport by wind field and turbulent diffusion). From the numerical point of view these four examples above can probably cover the skeleton of all reaction-diffusion problems. The corresponding equations, initial and boundary conditions with parameters used can be found in Table 1.

**4. Results and discussion**

    **4.1 Reaction-diffusion problems**

The first application is a simple diffusion problem without any reactions (Table 1). Here a chemically inert compound diffused from the centre of the computational domain. During this process the diffusing species can reach outer regions (Figure 2). The simulation of this problem is essentially calculating the Laplacian operator on all of the grid points and updating concentrations every time step. Therefore, it is easier to review the computational solution and performance in this simple case before the main concepts can be applied to more complex simulations.

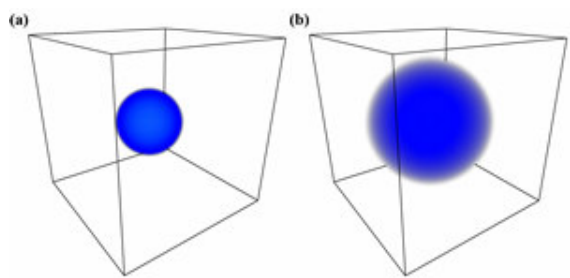

**Figure 2** Diffusion structure of a chemical species from the centre at (a) $t = 5\times10^3$ and (b) $t = 5\times10^4$. A detailed parameter set used in the simulation can be found in the Table 1.

The main question in computing the Laplacian is how to split the computational job on the domain (grid) to smaller rectangular blocks, which will be assigned to thread blocks, where a single thread updates a single grid point. Using the stencil in Eq. (4), data from nineteen grid points should be read to update one point, but since these are neighbouring points, data of every point will be read nineteen times (by its thread and its neighbours) while updating all of the grid. In order to avoid reading so many times from the slow global memory, data should be copied to the shared memory of a thread block. Using this stencil, a rectangular block will require an extra layer of grid points around it to allow computation on all points in the block. To read this data, one way is that after all threads read their corresponding grid points into shared memory, some threads read the extra layer. The other way is that although all threads read their corresponding grid points to shared memory, the ones on the edge do not compute, and the blocks overlap in every direction. We found that the second approach is generally faster. We can also utilize the fact that block indices are only two dimensional in CUDA (though blocks themselves are three-dimensional). We cannot assign all the "blocks" of grid points to thread blocks at once, instead, we assign only a thin, one block wide layer, and the kernels iterate though the simulated space in the third dimension. Since the blocks must overlap, the last two *z*-layers of data can be kept in the shared memory instead of reading them again from global memory, as the blocks iterate though the space (Figure 3). We call this solution the simple "Shared" method because it utilizes the shared memory in a simple way. It is very similar to the 3D finite difference computation example in the CUDA SDK [37], but in our case the stencil uses off-axis elements, so we need shared memory for all *z*-layers in the block, not only for the central *z*-layer. Moreover, the Shared method fits only the second generation of CUDA devices, because all global memory accesses are uncoalesced according to the strict rules of the first generation of CUDA devices.

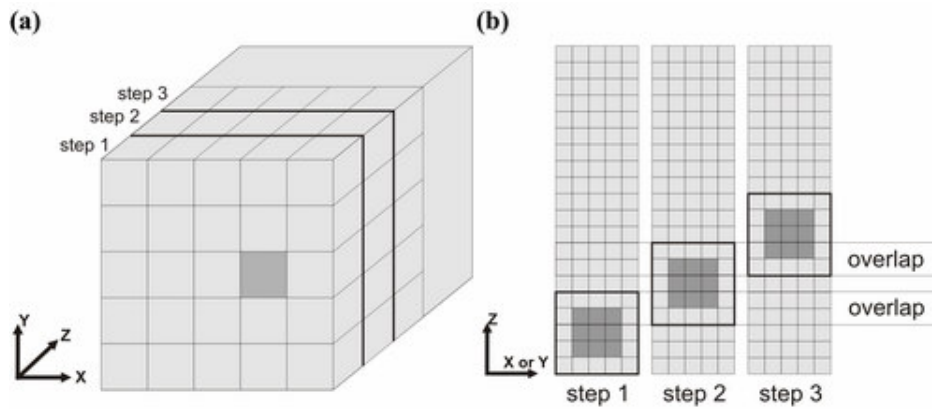

**Figure 3** Representation of the simulated space (a), cells indicate how the space is split and assigned to blocks in the Shared method. The first layer of points is assigned to blocks in the first step, then the second layer in the second step, etc. The darker block's evolution is depicted in (b). Here, each cell corresponds to a grid point in simulation. The black rectangle indicates data which is used to compute Laplacian in the points shown in darker color. The three columns are the same data, only shown separately to visualize iteration steps.

For first generation devices the thread blocks must iterate on the first dimension, and they must read and write global memory starting at an address multiple of 64 bytes, therefore the block width must be 16 (using 4-byte floats), to achieve coalesced access. Each block must use a wide tile of shared memory for $(1 + 2 \times 16) \times height \times depth$ elements. The extra shared memory is used as a streaming buffer for the global memory as depicted in Figure 4. In an iteration step data from the global memory is read to the second 16-element wide part of the shared memory. Then Laplacian can be computed on all elements of the first 16-element wide part, because the extra layer of data required on the left edge is provided from the previous step in the 1-element wide part, and the right edge is provided by the newly read data in the second 16-element wide part. After computing and writing the results back to global memory, the tile is moved to the right by 16 elements, and the iteration continues. The Moving Tiles method achieves coalesced memory access and it can be implemented without shared memory bank conflicts, because the block width is 16, which is equal to the number of shared memory banks. However, because of the high shared memory requirement less blocks can be allocated on a multiprocessor at once, limiting the performance for more complicated reaction-diffusion systems.

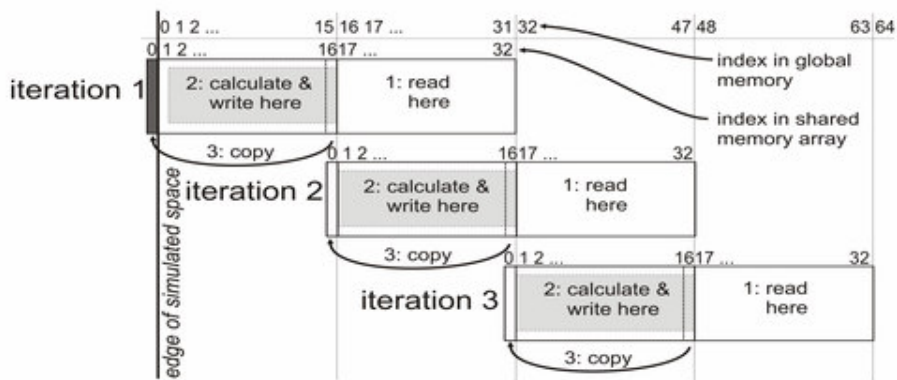

**Figure 4** Block iteration technique for approximation of the Laplacian in the Moving Tiles method. Coalesced reading and writing of global memory is achieved on the first generation of CUDA devices.

On the edge of the simulated space, the outermost grid points cannot be updated (the Laplacian cannot be computed), because they have no outer neighbours. Instead, these points are boundary points, their values are set according to the boundary conditions of the PDE before each time step. We use three separate kernels for updating these values on the left and right, top and bottom, and front and back sides of the simulated space, because the space is not necessarily cube shaped. Three additional kernels are updating the edges of the space parallel to the $x$, $y$, and $z$ axes. Corners do not need values because the stencil does not use them. These kernels are not optimized because their job is very small compared to the Laplacian computation, they contribute approximately 2% to the computational time.

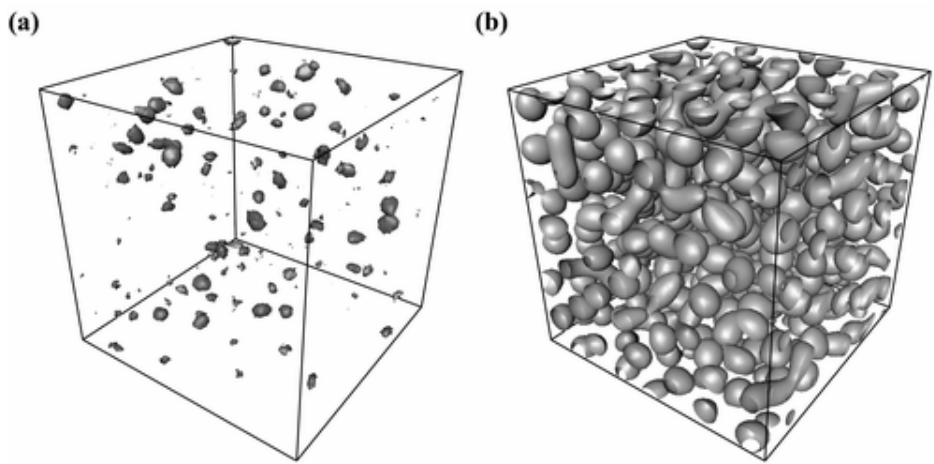

**Figure 5** Evolution of Turing patterns at (a) $t = 2\times10^5$ and (b) $t = 4\times10^5$. Domains in (a) and (b) represent the isosurface of $c_1 = 0.12$ and $c_1 = 0.05$, respectively. A detailed parameter set used in the simulation can be found in the Table 1.

Our second simulation example is a very extensively studied problem, both theoretically [38–40] and experimentally [4,5,41], in reaction-diffusion systems. Turing pattern formation occurs in case of sustained nonequilibrium conditions, where spatial patterns arise from an instability in a uniform medium. It is believed that several pattern formations in biology could be described by similar models (e.g. skin of certain animals) [42]. From the mathematical point of view the simplest one is a two-variable so-called "activator-inhibitor" model (Table 1). The activator generates itself by an autocatalytic process and also activates the inhibitor.

However, inhibitor can disrupt this autocatalytic process. The necessary condition for Turing pattern formation is that the diffusion coefficients of the activator and inhibitor species should be different. The simulation is started from the homogeneous distribution of the both species introducing small perturbation in initial conditions. During the evolution the effect of these small spatial perturbations was more and more pronounced regarding visual appearance of pattern via this specific reaction-diffusion mechanism (Figure 5).

Implementation of this problem is very similar to the solution of the diffusion equation. There are two species and corresponding arrays in global memory instead of one. In the kernel, twice as many shared memory is required. Again, every computational step of the Laplacian should be done twice, once for each species. These computations cannot be separated because of reaction terms in the equations, which make them *coupled* PDEs. If we calculated the diffusion and the reaction separately then we would have to read and write the data of every species at least twice.

The third model presents the pattern formation through a phase separation in the wake of a moving diffusion front [43,44]. Cahn–Hilliard equation was used to describe the phase separation [45]. This equation numerically is very challenging, because it contains fourth order derivatives. There are three processes included. First, the reaction of two electrolytes (which were initially separated in space) yields a chemical compound called intermediate species. This reaction provides the source for the precipitation, which is modeled as a phase separation of this intermediate product described by the Cahn–Hilliard equation with a source term. Finally, precipitate can be redissolved by the excess of one of the initial electrolytes, and this appears as a sink term both in the Cahn–Hilliard equation and in the reaction-diffusion equation (Table 1). Detailed experimental and theoretical description of this phenomenon can be found in Refs [43, 46–48]. During the evolution of the pattern, first a homogeneous precipitation layer forms, which travels through the medium via coarsening of the pattern (Figure 6).

During the computation data (concentrations) of all species must be loaded to shared memory at the same time, because all equations are coupled to each other. However, the Cahn–Hilliard equation is a fourth order PDE, it contains a biharmonic operator. Approximation of this operator is usually a non-compact stencil (using second neighbours as well as first neighbours). However, the stencil itself is numerically the same as applying the stencil of Laplacian twice on the data. Therefore, the biharmonic operator can be approximated by computing the Laplacian, applying boundary conditions again then computing Laplacian again on this data. The first Laplacian computation must be separate and completely finished before the rest of the computation is started, therefore two computing kernels were used. The first one is essentially the same as the kernel for diffusion, only without time integration. The second kernel must read data from arrays of all species (A, B and C) and Laplacian of the intermediate product (C), compute all reaction and diffusion terms then update concentrations of A ($c_1$), B ($c_2$) and C ($c_3$). This is a large kernel that requires many resources to be launched and have a significant computational time.

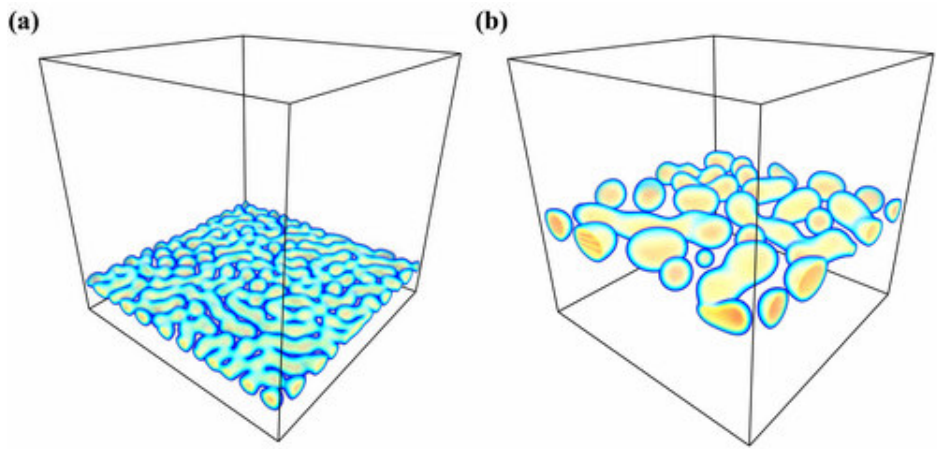

**Figure 6** Evolution of a moving precipitation pattern at (a) $t = 8\times10^3$ and (b) $t = 9\times10^4$. Pattern moves upward. A detailed parameter set used in the simulation can be found in the Table 1.

Our last example is an advection-diffusion problem, which has a great relevance in air pollution modelling. Solving diffusion-advection equations to describe the spread and/or transformation of air pollutants is a very important computational and environmental task (Table 1) [49,50]. The numerical simulations must be obviously achieved faster than in real time in order to use them in decision support [51]. A feasible way is the parallelization of the source code using supercomputers, clusters or GRID [52–56]. However, only a few preliminary trials have been presented to use GPU computing for air quality modelling [24–26]. Figure 7 shows the structure of the plume of an inert species originated from a single point source with a sinusoidal advection field.

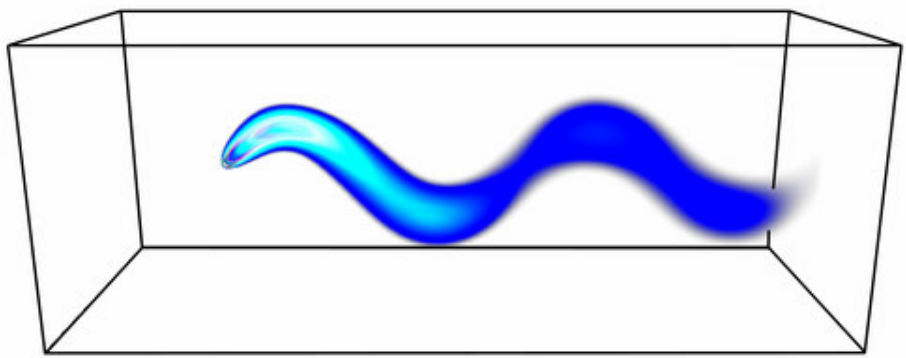

**Figure 7** Plume structure of an inert chemical species in the atmosphere originated from a point source. Red and black colour correspond to the high and low concentration of air pollutant. The size of the domain is 144 × 144 × 384. A detailed parameter set used in the simulation can be found in the Table 1.

From the computational point of view this simulation is very similar to the simple diffusion problem, however, an extra advection term is added (Table 1). Data read into the shared memory for Laplacian computation can be used to approximate first derivatives for advection. An upwind approximation was used to provide a stable solution.

A sample source code for both CPU and GPU versions are freely available to download from a web page [54], terms of use are also included on this page.

**4.2 Performance**

All performance tests were carried out on a desktop computer with 2.8 GHz Core 2 Duo processor, 3.0 GB RAM, and 32-bit Windows operating system. The reference CPU implementations of the simulations use a single thread on the CPU. They are compiled from a single source file to allow inline expansion for all functions by the compiler, which allows for about 20% speedup. We believe that in order to have a fair comparison of CPU and GPU computational speeds, both versions should be optimized, not just only the GPU version. The GPU tests were performed with two video cards: a GeForce 8800 GTX, which represents the first generation of CUDA-capable devices, and a GeForce GTX 275, which belongs to the second generation. Both the Shared method and the Moving Tiles method were tested for all simulations on both video cards, on a space of 192×192×192 grid points. Execution times were measured using timer functions provided by CUDA and averaged over thousands of time steps. The results are shown in Table 2 and Figure 8. Numerical parameters for the simulations can be found in Table 1, pseudocode for the reference CPU version and the kernels of both GPU solutions are in Table 3, 4 and 5, respectively.

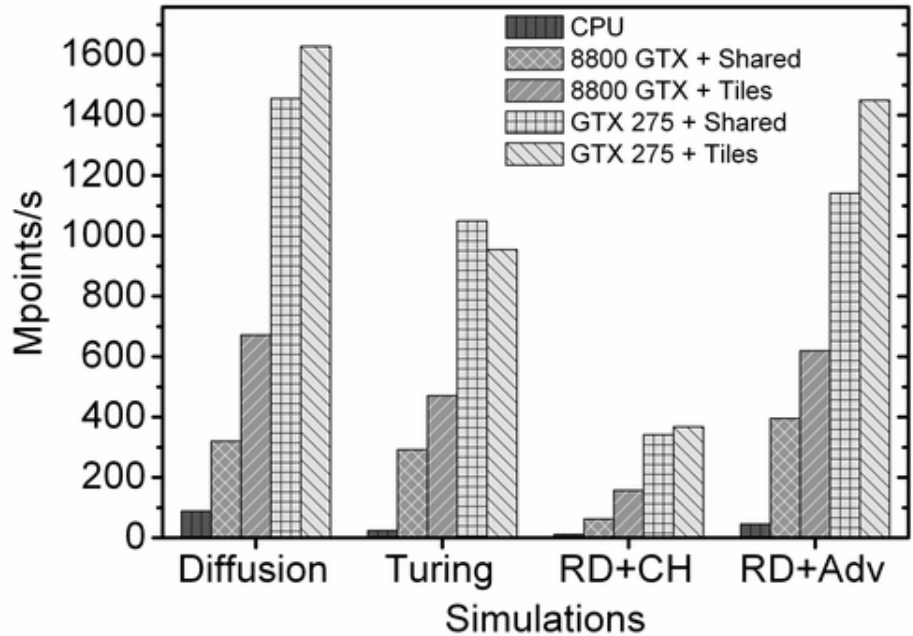

**Figure 8** Performance analysis on a first (GeForce 8800 GTX) and a second (GeForce GTX 275) generation video cards solving four different reaction-diffusion problems with two parallelization strategies.

The simulation speed were measured by the number of grid points where the simulation can be advanced with one time step, in one second, which gives the Mpoints/s unit. The figures indicate effective simulation speed, they also include the computation of boundary conditions. However, this unit cannot be used to compare different simulations, only implementations of the same simulation problem. Generally, the more complicated simulation provides the lower Mpoints/s values.

According to the measurements, the Moving Tiles method is almost always faster than the Shared method. On the 8800 GTX its relative advantage is big, because the Shared method causes uncoalesced memory access on the first generation CUDA devices. On the GTX 275 the Moving Tiles is still faster (except for the Turing simulation), but the relative advantage is smaller. This is because although all memory operations are coalesced on the second generation devices, the unaligned memory requests of the Shared method are usually serviced in two coalesced memory operations, but the aligned requests of the Moving Tiles method are always serviced in one coalesced memory operation.

The GTX 275 is much faster than the 8800 GTX in every simulation and in both methods, because it is a more recent video card and it has 240 streaming processors while the other one has 128.

**5. Conclusion**

We presented in this study a potential application of GPUs to solve reaction-diffusion equations. These equations arise in numerous scientific areas and are responsible to describe patterns and structures of involved chemical species. Moreover, using a similar framework, the air pollution modelling can be simulated using this new parallel infrastructure. Diversified systems have been tested to present the efficiency of GPU computing. We can conclude that the parallel implementations achieve typical acceleration values in the order of 5–40 times compared to CPU using a single-threaded implementation on a 2.8 GHz desktop computer depending on the problem and parallelization strategy used. Our results indicate that the GPU computing would be a promising and cost efficient tool to run parallel applications to solve reaction-diffusion and air quality problems.

**Acknowledgement**

The authors thank Prof. Zoltán Rácz (Eötvös University) for many helpful discussions. Authors acknowledge the financial support of the Hungarian Research Found (OTKA K68253 and K81933). This work makes use of results produced by the SEE-GRID eInfrastructure for regional eScience, a project co-funded by the European Commission (under contract number 211338) through the Seventh Framework Program. SEE-GRID-SCI stimulates widespread

eInfrastructure uptake by new user groups extending over the region of South Eastern Europe, fostering collaboration and providing advanced capabilities to more researchers, with an emphasis on strategic groups in seismology, meteorology and environmental protection. Full information is available at http://www.see-grid-sci.eu

## Tables

**Table 1** Reaction-diffusion problems with equations, parameters, initial and boundary conditions used in this study.

| Phenomenon | Equations | Parameters | Initial conditions | Boundary conditions |
|---|---|---|---|---|
| Pure diffusion | $\frac{\partial c}{\partial t} = D\nabla^2 c$ | $D = 1$, $h = 1$, $\delta t = 0.02$ | $c_0 = 1.0$ inside a sphere ($r = 20$) in the middle, $c_0 = 0.0$ elsewhere | no-flux |
| Turing pattern formation | $\frac{\partial c_1}{\partial t} = D_1 \nabla^2 c_1 + c_1 - c_1^3 - c_2$, $\frac{\partial c_2}{\partial t} = D_2 \nabla^2 c_2 + \gamma(c_1 - \alpha c_2 - \beta)$ | $D_1 = 5.0\times 10^{-5}$, $D_2 = 5.0\times 10^{-3}$, $h = 6.2\times 10^{-3}$, $\delta t = 5.0\times 10^{-4}$, $\alpha = 0.5$, $\beta = 0.09$, $\gamma = 26.0$ | $c_{10} = 1.0 + \sigma$, $c_{20} = 1.0 + \sigma$, $\sigma$ = uniform random between $-5.0\times 10^{-4}$ and $5.0\times 10^{-4}$. | no-flux |
| Phase separation behind a chemical front using Cahn–Hilliard equation | $\frac{\partial c_1}{\partial t} = D_1 \nabla^2 c_1 - k_1 c_1 c_2 - k_2 c_1 c_3$, $\frac{\partial c_2}{\partial t} = D_2 \nabla^2 c_2 - k_1 c_1 c_2$, $\frac{\partial c_3}{\partial t} = k_1 c_1 c_2 - k_2 c_1 c_3 - \lambda \nabla^2 \left( \varepsilon c_3 - \gamma c_3^3 + \sigma \nabla^2 c_3 \right)$ | $D_1 = D_2 = 1$, $h = 1$, $\delta t = 0.02$, $k_1 = 0.2$, $k_2 = 0.005$, $\lambda = \varepsilon = \gamma = \sigma = 1.0$ | $c_{10} = 0.0$, $c_{20} = 1.0$, $c_{30} = -1.0$ | no-flux for $c_2$ and $c_3$, Dirichlet for $c_1 = 10.0$ at the $x = 0$ plane, noflux for $c_1$ at other sides of simulated space |
| Atmospheric advection-diffusion process | $\frac{\partial c}{\partial t} = -\nabla \cdot (\vec{u} c) + D\nabla^2 c + E$ | $D = 100$ m$^2$ s$^{-1}$, $h = 100$ m, $\delta t = 5$ s, $u_x = 5$ m s$^{-1}$, $u_y = 1.0$ m s$^{-1}$, $u_z = 5.0\ \sin(t/500\text{s})$ m s$^{-1}$, $E = 10$ mol dm$^{-3}$ s$^{-1}$ (emission term) | $c_0 = 0.0$ | no-flux |

**Table 2** The relative speedup using a first (GeForce 8800 GTX) and a second (GeForce GTX 275) generation video cards compared to a single-threaded implementation on a 2.8 GHz desktop computer.

| System | GeForce 8800 GTX | | GeForce GTX 275 | |
| --- | --- | --- | --- | --- |
| | Shared | Tiles | Shared | Tiles |
| Diffusion | 3.6 | 7.6 | 16.4 | 18.3 |
| Turing pattern | 12.0 | 19.4 | 43.2 | 39.3 |
| Phase separation behind a chemical front using Canh–Hilliard equation | 5.6 | 14.2 | 30.8 | 33.2 |
| Atmospheric diffusion-advection | 8.6 | 13.5 | 24.9 | 31.6 |

**Table 3** Pseudocode for the CPU implementation. For GPU implementations, loops on line 4 and 6 are replaced by kernels, see Tables 4 and 5.

```
1   allocate two float arrays for each species: first, second
2   load initial values to first arrays
3   for step = 1 to maxSteps
4      for all boundary points
5          calculate boundary condition in first arrays
6      for all grid points
7          compute Laplacian term from first arrays
8          compute reaction terms from first arrays
9          store updated values in second arrays
10     swap first and second arrays
11     export first arrays if necessary
12  free allocated memory
```

**Table 4** Pseudocode for the kernel of the Shared method. Variables tx, ty and tz represent the current thread's x, y and z indices, variables bW, bH and bD represent the block's width, height and depth (the number of threads inside it), respectively. The code indicates input and output arrays only for one species, for simplicity. In real programs, they represent separate arrays for all species involved in the simulation.

```
kernel SharedMethod( float array C_in[Depth][Height][Width],
             float array C_out[Depth][Height][Width])
   declate shared float array Cs[bD][bH][bW]
   i = index for C_in[tz][by*(bH-2)+ty][bx*(bW-2)+tx]
   read C_in[i] into Cs[tz][ty][tx]
   for k = 0 to Depth/(bD - 2)
      synchronize threads
      if this thread is not on the edge of the block
         compute Laplacian from Cs array
         compute reaction and/or advection terms from Cs array
         store updated value in C_out[i]
      synchronize threads
      if tz >= bD-2
         Cs[tz-bD+2][ty][tx] = Cs[tz][ty][tx]
      synchronize threads
      i = index for C_in[(k+1)*(bD - 2)][by*(bH-2)+ty][bx*(bW-2)+tx]
      if tz >= 2
         read C_in[i] into Cs[tz][ty][tx]
   end for
end kernel
```

**Table 5** Pseudocode for the kernel of the Moving Tiles method. Variables tx, ty and tz represent the current thread's x, y and z indices, variables bW, bH and bD represent the block's width, height and depth (the number of threads inside it), respectively. The code indicates input and output arrays only for one species, for simplicity. In real programs, they represent separate arrays for all species involved in the simulation.

```
kernel MovingTiles(  float array C_in[Depth][Height][Width],
            float array C_out[Depth][Height][Width])
   declate shared float array Cs[bD][bH][2*bW+1]
   i = index for C_in[by*(bD-2)+tz][bx*(bH-2)+ty][tx]
   read C_in[i] into Cs[tz][ty][tx+1]
   i = index for C_in[by*(bD-2)+tz][bx*(bH-2)+ty][bW+tx]
   read C_in[i] into Cs[tz][ty][bW+tx+1]
   for k = 0 to Width / bW
      i = index for C_in[by*(bD-2)+tz][bx*(bH-2)+ty][k*bW+tx]
      synchronize threads
      if this thread is not on the edge of the block
         compute laplacian from Cs array
             //data for thread(tz, ty, tx) is in Cs[tz][ty][tx+1]
         compute reaction and/or advection terms from Cs array
         store updated value in C_out[i]
      synchronize threads
      if tx == bW-1
         Cs[tz][ty][0] = Cs[tz][ty][bW]
      Cs[tz][ty][tx+1] = Cs[tz][ty][bW+tx+1]
      synchronize threads
      if k <= Width/bW − 1
         i = index for C_in[by*(bD-2)+tz][bx*(bH-2)+ty][(k+2)*bW+tx]
         read C_in[i] into Cs[tz][ty][bW+tx+1]
   end for
end kernel
```